\documentclass[journal]{IEEEtran}   
\usepackage[T1]{fontenc}  
\usepackage[english,italian,french]{babel}
\usepackage{amsmath,amsfonts,amssymb,bm,mathrsfs,upgreek} 
\usepackage{graphicx,color}
\usepackage[dvipsnames]{xcolor}
\IEEEoverridecommandlockouts
\usepackage[final]{changes} 

\newcommand{\micron}{\ensuremath{\upmu\mathrm{m}}}    
\newcommand{\micros}{\ensuremath{\upmu\mathrm{s}}}    
\newcommand{\microW}{\ensuremath{\upmu\mathrm{W}}}    
\newcommand{\Ohm}{\ensuremath\Omega}                  
\newcommand{\unit}[1]{\ensuremath{\mathrm{#1}}}       

\graphicspath{{./Figures/}}

\begin{document}
\selectlanguage{english}
\bstctlcite{IEEEexample:BSTcontrol}

\title{Phase noise measurement\\of semiconductor optical amplifiers}
\author{Damien Teyssieux, Martin Callejo, Jacques Millo, Enrico Rubiola, and Rodolphe Boudot
  \thanks{The authors are with Institut FEMTO-ST, CNRS, Université Marie et Louis Pasteur, ENSMM, 26 chemin de l'Épitaphe, 25030 Besançon Cedex, France.}
  \thanks{Enrico Rubiola is also with INRIM, Strada delle Cacce 91, 10135 Torino, Italy.}
  \thanks{Corresponding author Damien Teyssieux, ORCID: 0000-0001-9964-0371, email damien.teyssieux@femto-st.fr,\\
    Martin Callejo, ORCID: 0000-0001-5294-7640,\\
    Enrico Rubiola, ORCID: 0000-0002-5364-1835, web page https://rubiola.org,\\
    Rodolphe Boudot, ORCID: 0000-0002-0933-2541.
  }
}
\maketitle

\begin{abstract}
\boldmath
We introduce a novel measurement method for the phase noise measurement of optical amplifiers, topologically similar to the Heterodyne Mach-Zehnder Interferometer but governed by different principles, and we report on the measurement of a fibered amplifier at 1.55~\micron\ wavelength.   The amplifier under test (DUT) is inserted in one arm of a symmetrical Mach-Zehnder interferometer, with an AOM in the other arm.  We measure the phase noise of the RF beat detected at the Mach-Zehnder output.  
The phase noise floor of the amplifier decreases proportionally to the reciprocal of the laser power at the amplifier input, down to \(-126\)~\unit{dBrad^2/Hz} at \(f=100\)~\unit{kHz}. 
The DUT flicker noise cannot be measured because it is lower than the background of the setup.  This sets an upper bound of the amplifier noise at \(-32\)~\unit{dBrad^2/Hz} at \(f=1\)~Hz, which corresponds to a frequency stability of \(5.2{\times}10^{-17}/\tau\) (Allan deviation), where \(\tau\) is the integration time.  Such noise level is lower than that of most Fabry-Pérot cavity-stabilized lasers. 
These results are of interest in a wide range of applications including metrology, instrumentation, optical communications, or fiber links.
\unboldmath
\end{abstract}

\section{Introduction}\label{introduction}
Optical amplifiers of different technologies (semiconductor amplifiers \cite{Connelly:2002, Dutta:2013}, Erbium-doped fiber amplifiers \cite{Desurvire:2002}, and Raman amplifiers \cite{Islam-2002}) are key devices of fiber optics communication and network systems, where they are needed to compensate for transmission losses of the fiber ($\sim$\,0.2~dB/km).
On our side, we are mostly interested in the generation and dissemination of ultra-stable and low-phase-noise optical signals for metrology.  Quantum clocks at optical frequencies, which exhibit frequency stability of \(10^{-18}\ldots10^{-19}\) \cite{Ludlow-2015, Takamoto-2020, Aeppli-2024}, are under consideration for the re-definition of the SI second \cite{Dimarcq-2024}.
Such clocks are compared to one another through local or long-distance two-way links compensated for the length fluctuation of the optical fiber \cite{Foreman-2007, Terra-2010, Stefani-2015, Clivati-2020, Cantin-2021, Beloy-2021}, which in turn rely on low-noise optical amplifiers. 

Modern \replaced{vapor-cell}{chip-scale} quantum clocks, based on sub-Doppler spectroscopy techniques and exhibiting short-term stability in the low \(10^{-13}\) at 1~s, are another domain where low-noise optical amplifiers are of paramount importance.  Among the various approaches, it is worth mentioning the two-photon spectroscopy of the Rb atom at 778 nm \cite{Nez-1993, Newman-2019, Newman-2021, Callejo-2025}.
The major contributions to their short-term instability are the photon shot noise and the intermodulation \added{effect} \cite{Audoin-1991} induced by the free-running frequency noise of the interrogation laser.

Phase noise is a well established topic in RF and microwaves (see IEEE Standard 1139 \cite{Donley:2022-IEEE-1139} and the tutorial/review article \cite{Rubiola-2023-Companion}).  Concepts and tools are general, and equally appropriate to optics.  We remind that phase noise should be reported as the power spectral density (PSD) of the random phase \(\varphi(t)\) in \unit{rad^2/Hz} (the appropriate SI unit) as a function of the Fourier frequency \(f\), and denoted with \(S_\varphi(f)\).  However, the non-SI quantity \(\mathscr{L}(f)\) is most often encountered, defined as \(\mathscr{L}(f)=(1/2)\,S_\varphi(f)\) and given in dBc/Hz \cite[Eq.\,(1) p.\,13 and related text]{Donley:2022-IEEE-1139}.

Phase noise in RF and microwave amplifiers has been studied extensively \cite{Walls-1997, Ferre-Pikal-2008, Cibiel-2004b}, and discussed in detail in \cite{Boudot-2012}.  Conversely, phase noise of optical amplifiers got little consideration. 
Kikuchi \cite{Kikuchi-1991} addresses the problem, but the experimental method is only suitable to large Fourier frequencies with poor frequency resolution (tens of MHz, inferred from \cite[Fig.\,4]{Kikuchi-1991}), likely because the self-heterodyne method was not used.
Other articles report on RIN \cite{Danion-2014} or on the phase noise of optically carried microwave signal \cite{Auroux.2014, Gordon-1990, Cahill-2017, Eliyahu-2008}, often in the context of fiber links.  
There are theories and simulations on phase noise which provide some experimental results \cite{Moro-2010,Shimizu-2024,Wei.2005}, but we find them unsatisfactory or confusing.  For example, \cite{Shimizu-2024} repeatedly gives the `PSD of phase noise' in \unit{Hz^2/Hz}, but such unit is appropriate to express a frequency noise, and \cite{Wei.2005} only gives the variance of the random phase.  In parallel to our study, NASA JPL investigated on semiconductor optical amplifiers at \({\approx}852\:\unit{nm}\) (D2 line of \unit{{}^{133}Cs} atom) to be used in a cooling system for quantum sensors \cite{Kittlaus-2025}.  Among other parameters ultimately intended for a future flyable system, such as switching speed, extinction ratio, environment sensitivity and power consumption, Ref.~\cite[Sec.\,2.5 `Added noise']{Kittlaus-2025} provides some data about phase noise with little technical details.

\added{Deeper reading of the literature reveals a common misconception.  Indeed optical amplifiers are generally tested by measuring the laser noise with and without the amplifier \cite[Fig.\,3a and Fig.\,6]{Kittlaus-2025}, \cite[Fig.\,1--2]{Terra-2010}, \cite[Fig.\,8--9]{Zhao-2023}.  There results an unnecessarily high background noise because such schemes do not take benefit from a differential measurement, which divides the input-output phase of the DUT from the laser's  low-frequency divergent phenomena (\(1/f^2, 1/f^3, 1/f^4\), etc.).}

In the following Sections, we introduce a measurement method \added{which fixes the problem mentioned,} \replaced{and we show}{with} the application to the phase noise of an optical amplifier.  Unlike the existing literature, we focus on the method, and we provide technical details of the experiments.

\begin{figure}[t]   
  {\centering\includegraphics[width=0.9\columnwidth]{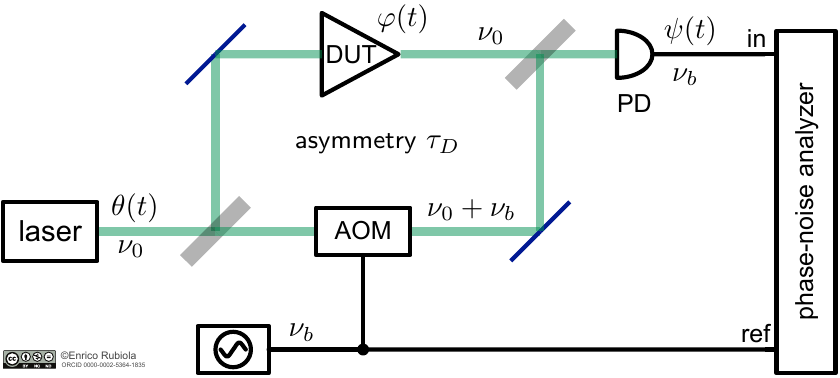}\par}
  \caption{Principle of the optical phase noise measurement.  AOM: Acousto-Optic Modulator, DUT: \replaced{D}{d}evice \replaced{U}{u}nder \replaced{T}{t}est (optical amplifier), PD: quantum \replaced{PhotoDetector}{photodetector}.}  
  \label{fig:MZ-interferometer}
\end{figure}

\begin{figure*}[t]   
  {\centering\includegraphics[width=0.9\textwidth]{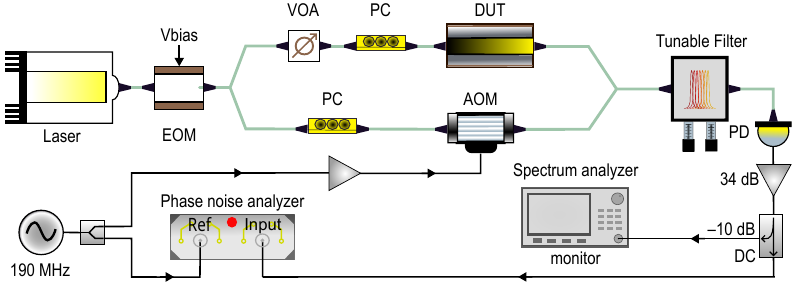}\par}
  \caption{Experimental setup used for optical phase noise measurement.  AOM: Acousto-Optic Modulator, DUT: \replaced{D}{d}evice \replaced{U}{u}nder \replaced{T}{t}est (semiconductor optical amplifier), EOM: Electro-Optic Modulator, PC: Polarization Controller, VOA: Variable Optical Attenuator, PD: Photodiode, DC: 10 dB Directional Coupler.}  
  \label{fig:SETUP}
\end{figure*}

\section{Measurement principle}\label{Sec:principle}
Our scheme (Fig.~\ref{fig:MZ-interferometer}) is a Mach-Zehnder (MZ) interferometer with the amplifier under test (DUT) is in one arm and an acousto-optic modulator (AOM) in the other arm to shift the optical frequency \(\nu_0\).  The beat between the two optical beams results in the radio-frequency \(\nu_b\) at the photodetector output.
The quantity \(\tau_D\) is the delay asymmetry between the two arms of the MZ, and the quantities \(\theta(t)\), \(\varphi(t)\) and \(\psi(t)\) are the random phases of the laser, of the DUT, and at the photodetector output, respectively.

Figure\,\ref{fig:MZ-interferometer} is topologically similar to other flavors of the heterodyne, or self-heterodyne MZ interferometer.
With large \added{delay} asymmetry \(\tau_D\), no DUT, and a spectrum analyzer instead of the phase-noise analyzer, the interferometer measures the laser frequency noise.  This is known as the delayed self-heterodyne interferometer (DSHI), generally attributed to Okoshi \cite{Okoshi-1980}, and widely used since \cite{Richter-1986, Zhao-2022, Ishida-1991, Li-2019, Peng-2021, Camatel-2008, Llopis-2011}.
In contrast, a symmetric MZ (\(\tau_D\approx0\)) \replaced{rejects the laser phase noise \(\theta(t)\).  Thus, the phase \(\psi(t)\) at the photodetector output is equal to the DUT random phase \(\varphi(t)\) transposed from optics to RF\@.}{does a differential phase measurement of the DUT phase, rejecting the laser frequency noise.}  \added{Using the RF oscillator as the reference of the phase noise analyzer, this scheme does a true differential measurement of the DUT phase noise, i.e., DUT input-to-output phase fluctuation \(\varphi(t)\).} 
This \added{idea} was used by Lavan \cite{Lavan-1976} to monitor the turbulence of a gas with an oscilloscope. 

The main difference from other schemes---that we use a phase noise analyzer instead of \added{an oscilloscope or} an RF spectrum analyzer---is far less trivial than it seems.  Firstly, the system works only with a modern fully-digital phase noise analyzer, which allows arbitrary phase relationship between inputs, and even track\added{ing} multiple cycles.  Conversely, classical analyzers are not suitable because the phase detector (a saturated double-balanced mixer) requires that the two inputs are in quadrature within a small fraction of a rad.  In optics, this is unrealistic.  Secondly, a phase-noise analyzer enables the measurement of noise sidebands at a level out of reach for a spectrum analyzer, and at smaller Fourier frequencies.  Finally, the measurement requires appropriately low noise and low instability in the optical part.  In fact, the DUT internal delay is small, and does not suffer from the large, diverging phase noise which affects \deleted{all} oscillators \added{and lasers} (Leeson effect, \cite[Chapters\,3--4]{Rubiola:2010}). 

Provisionally restricting our attention to the laser and to the DUT noise, the phase noise at the photodetector output is 
\begin{align}
  S_\psi &= \left|H(f;\tau_D)\right|^2\,S_\theta(f) + S_\varphi(f)
  \label{eqn:Spsi}
\end{align}
where \(S(f)\) is the PSD of the quantity in the subscript, and \(|H(f;\tau_D)|^2\) is the transfer function 
\begin{align}
  \left|H(f)\right|^2 
  &=4\sin^2(\pi\tau_D f)
  \label{eqn:H2f}\\
  &\simeq 4\pi^2\tau_D^2 f^2~~~\text{for \(\pi\tau_D f\ll1\)}
  \label{eqn:H2f2}
  \end{align}
which results from the delay imbalance \(\tau_D\).
The proof is given in \cite[Sec.~2C, `Delay Line Theory']{Volyanskiy-2008}.  \deleted{Interestingly, this proof does not depend on the nature of the modulation, thus it holds for the fractional amplitude noise \(S_\alpha(f)\) as well.}

Equation \eqref{eqn:H2f} is most often used with large \(\tau_D\) and no DUT (\(\varphi=0\)) to calculate the laser noise \(S_\theta(f)\) from \(S_\psi(f)\) measured at the output.  Notice that the maximum frequency is limited to \(f<1/\pi\tau_D\).  Beyond, \(|H(f;\tau_D)|^2\) has a series of periodic zeros where \eqref{eqn:Spsi} cannot be inverted.  In contrast, we use \eqref{eqn:Spsi}-\eqref{eqn:H2f} to determine the maximum imbalance \(\tau_D\) that can be tolerated for \(S_\theta(f)\) to be rejected to a level lower than the DUT noise \(S_\varphi(f)\).

\section{Experimental setup}
The experimental setup is shown in Fig.~\ref{fig:SETUP}, where all the optical parts are fibered.  

The DUT (Thorlabs BOA1150P) is a C-band (1.55~\micron) optical amplifier in InP/InGaAsP multiple quantum well technology.  It exhibits 27~dB small-signal gain, 18~dBm saturated output power, 8.5~dB nominal noise figure, and 105~nm bandwidth (\(-3\)~dB).  It is available as a standard 14-pin butterfly package with FC/APC connectors.  Polarization-maintaining fibers (PM15-U40A) are used at both input and output.  A temperature control provides gain and spectrum stabilization, relying on an integrated TEC thermistor. 

The laser source is a low-noise Koheras Adjustik.  
The EOM (iXblue MXAN-LN-20, $V_{\pi}$~$\simeq$~5.5~V) at the output of the laser is used to adjust the optical power without changing the laser operation point.  It is biased by a commercial voltage supply (Keysight E3620A).  There is no need of temperature control because the EOM is stable enough for the duration of the measurement (2-10 minutes).  \added{Because there is no point in having electrical control on power, the EOM can be replaced with a variable optical attenuator at will.}

The optical power splitter and combiner, and the optical parts in between, form the MZ interferometer.  The DUT is preceded by an optical attenuator (VOA, IDIL COCOM03898) which enables to set the DUT power as desired.  The Acousto-Optic Modulator (AOM\added{, Gooch \& Housego `Fiber-Q' SFO2019-T-M200-0.1C2J-3-F2S-04}) in the other arm is driven by a 190~MHz RF signal from a commercial frequency synthesizer \added{(Rodhe Schwarz SMB100A)}. 
The photodiode (PD) is a home-made assembly consisting of a Thorlabs FGA01 (InGaAs technology, 1~GHz bandwidth), followed by an RF amplifier.  It provides the 190~MHz beat note, which is further amplified. 

A narrowband filter at the photodetector input is necessary to control the amplified spontaneous emission (ASE) of the amplifier, which is a wideband optical process.  Without such filter, the ASE results in unnecessarily high white noise floor at the photodetector output, and under some experimental conditions it spoils the phase-noise sensitivity.  We used a Yenista WSM-160 filter, tunable in center frequency (1.525--1.570 \micron) and bandwidth (0.25--45 nm), set at the narrowest bandwidth.

The phase noise analyzer (Microchip 53100A) compares the 190~MHz beat note to the reference, and shows the phase noise PSD at Fourier frequencies \(f\) from 1 Hz to 1 MHz.  This instrument is based on direct digitization of the input signals, I-Q detection, and correlation-and-averaging algorithm to reduce the instrument background noise \cite[Sec.~2.5]{Rohde:2021}.

All the experiments are done in a shielded chamber with PID control of temperature (\(22\pm0.5~\unit{^\circ C}\)) and humidity (\(50\%\pm10\%\)).
In addition, the setup has been thermalized for several days before measuring, and protected against air flow by a thermal blanket (survival kit).

\section{Background noise model}
By inspection on Fig.~\ref{fig:SETUP}, the phase noise $S_{\psi}(f)$ seen by the phase noise analyzer is
\begin{equation}
  S_{\psi}(f) = S_\text{AOM}(f) + S_\text{PD}(f) + S_\text{Amp}(f)
  \label{eqn:PN_lin2}
\end{equation}
where \(S_\text{AOM}(f)\), \(S_\text{PD}(f)\), and \(S_\text{Amp}(f)\) are the phase noise of the AOM, the phase detector, and the RF amplifier at the detector output, respectively.  The laser contribution \(S_\theta(f)\) is made negligible with \(\tau_d\approx0\) in \eqref{eqn:H2f}.
The low-noise 190 MHz RF signal is common mode\added{ for the RF measurement}, thus its fluctuations are rejected.  \added{As a matter of fact, if a phase \(\gamma(t)\) is added to such RF signal, the same phase \(\gamma(t)\) is present at the photodetector output and at the `ref' input of the phase-noise analyzer.}   Experience suggests that the contribution of the AOM and of its RF amplifier is negligible in our conditions.
The DUT noise \(S_\varphi(f)\) is not included \added{in \eqref{eqn:PN_lin2}} because this the measurand, not the background noise of the system.

Hereinafter, we restrict our attention to white phase noise.  
Physical insight suggests that the optical power at the photodetector input is
\begin{align}
  P(t) &= P_1 + P_2 + 2\sqrt{P_1P_2}\cos[2\pi\nu_bt]
\end{align}
where \(P_1\) and \(P_2\) is the power from the two arms.  The anti-ASE bandpass filter is narrow enough to make \(P_\text{ASE}\) negligible.

The photocurrent is
\begin{align}
  I(t) &= \rho\left(P_1 + P_2\right) + 2\rho\sqrt{P_1P_2}\cos[2\pi\nu_bt]  \label{eqn:PhotoI}
\end{align}
where \(\rho=q\eta/h\nu\) is the detector responsivity.  In turn, \(q=1.602{\times}10^{-19}~\unit{C}\) is the charge of the electron, \(\eta\) is the quantum efficiency, \(h=6.63{\times}10^{-34}~\unit{J/Hz}\) is the Planck constant, and \(h\nu\simeq1.28{\times}10^{-19}~\unit{J}\) is the photon energy.  To this extent, \(\nu_0\) and \(\nu_0+\nu_b\) are equivalent.  It holds that \(\rho\simeq1\)~A/W at 1.55~\micron\ with \(\eta=0.8\).

The photodiode has an internal termination \(R_0=50~\Ohm\).  The amplifier input is AC coupled, with \(R_0=50~\Ohm\) input impedance.  Thus, the DC flows in the diode internal resistor only, while RF and noise current are equally split between the two resistors.
The RF power at the amplifier input is 
\begin{align}
  P_\text{RF} &= \frac{1}{2} \rho^2 R_0P_1P_2
  \label{eqn:PRF}
\end{align}

The shot noise PSD is \(S_I=2qI_\text{DC}\), where \(I_\text{DC}\) is the DC part of \eqref{eqn:PhotoI}, that is, \(I_\text{DC}=\rho(P_1+P_2)\).  Because the load impedance seen by the shot current is \(R_0/2\), and only half power goes in the amplifier input, the PSD is
\begin{align}
  S_\text{sh} &= \frac{1}{2} R_0 q \rho\left(P_1+P_2\right)
\end{align}

The thermal-noise model is the textbook case of an amplifier of input impedance \(R_0\) and noise factor \(F\) input-terminated to a resistor \(R_0\) (the photodiode internal resistor) at temperature \(T\)\@.  Thus,
\begin{align}
  S_\text{th} &= FkT
\end{align}
where \(k=1.38{\times}10^{-23}~\unit{J/K}\) is the Boltzmann constant.

The white phase noise background is the white noise PSD divided by the carrier power, i.e.,
\begin{align}
  S_{\psi\,\text{bg}}(f) 
  &= \frac{S_\text{th}+S_\text{sh}}{P_\text{RF}} \qquad\qquad\text{(background)}\\[0.5ex]
  &= \frac{2FkT}{\rho^2 R_0P_1P_2} + \frac{q(P_1+P_2)}{\rho P_1P_2}
  \label{eqn:PNbg1}
\end{align}

The lowest background noise is achieved with \(P_1=P_2\).  Replacing \(P_1\) and \(P_2\) with \(P_0\), we get
\begin{align}
  S_{\psi\,\text{bg}}(f) &= \frac{2FkT}{\rho^2 R_0P_0^2} + \frac{2q}{\rho P_0}
  \label{eqn:PNbg2}
\end{align}
This equation identifies the threshold power  
\begin{align}
  P_\text{thr,\:opt} = \frac{FkT}{q\rho R_0}
  \qquad\text{(optical)}
\end{align}
which divides the thermal regime \(S_{\psi\,\text{th}}(f)\propto 1/P_0^2\) at low \(P_0\), from the shot regime \(S_{\psi\,\text{sh}}(f)\propto1/P_0\) at high \(P_0\).  For reference, with \(\rho=1~\unit{A/W}\) and \(F=2\) (3 dB noise figure) at room temperature, such threshold is 1~mW optical power at the photodiode input, thus 25~\microW\ RF power at the amplifier input.

\begin{figure*}[t]   
  {\centering\includegraphics[width=0.98\textwidth]{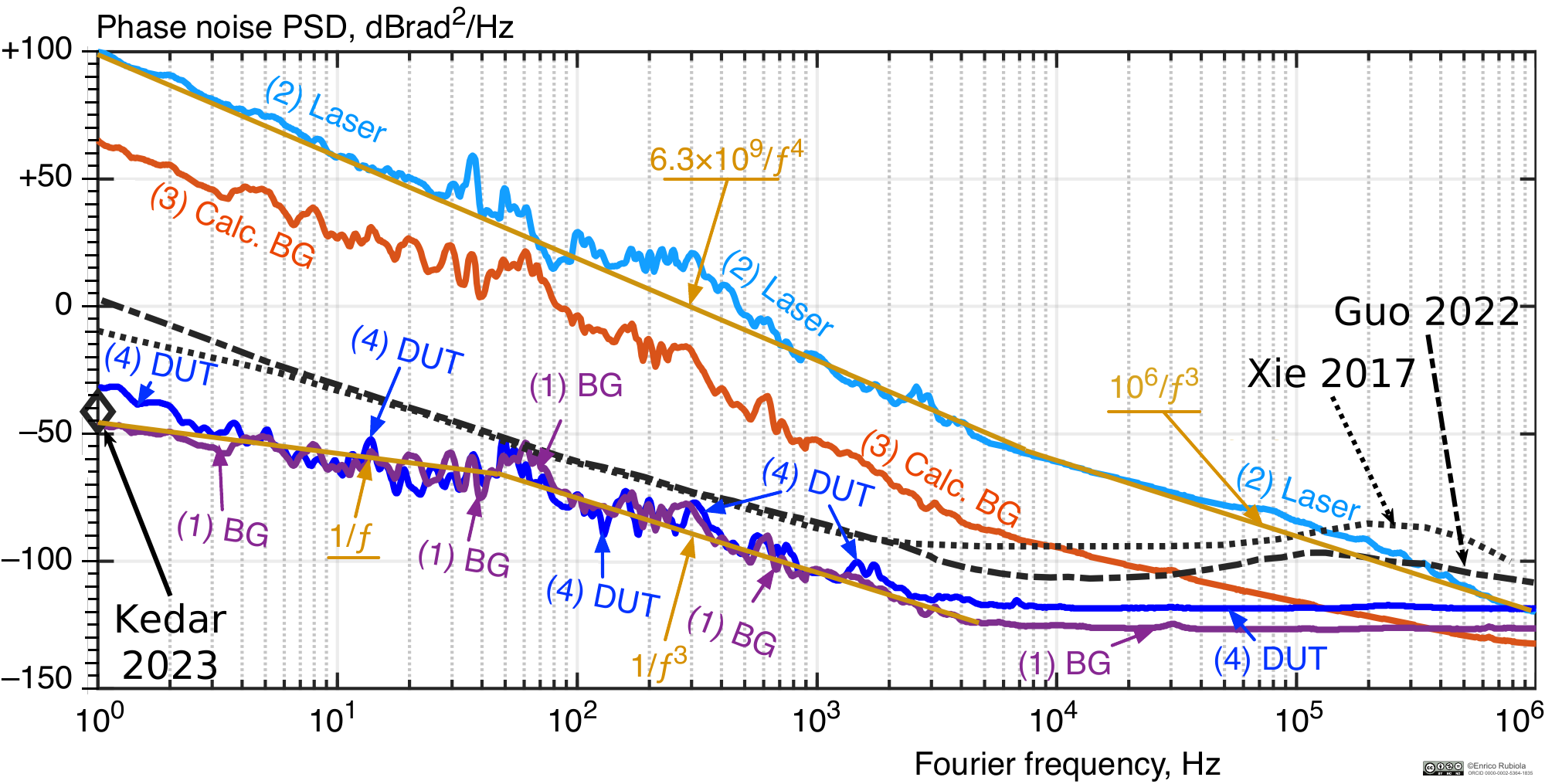}\par}
  \vspace*{-1ex}
  \caption{\replaced{Phase noise, as described in Sec.~\ref{sec:Experiment1}, measured with \(-5\)~dBm RF power at the photodiode output.  Curve 1: background noise, measured with no DUT, and negligible \(\tau_D\).  Curve 2: laser noise, measured with no DUT, and \(\tau_D=460\)~ns.  Curve 3: background in the measurement of the laser noise, calculated from curve 1 accounting for \(\tau_D=460\)~ns.  Curve 4: Total noise, DUT plus background, measured with negligible \(\tau_D\).}{Phase noise of the laser, background noise of the setup (with no DUT), and total noise (DUT plus background).  The RF power at the photodiode output is \(-5\)~dBm.}  The dotted and dash-dot \replaced{black}{gray} lines are the phase noise of high-performance cavity-stabilized lasers \cite{Guo-2022,Xie-2017b}.  The diamond symbol is the phase noise (\(S_{\varphi}(1\:\unit{Hz})=-42\:\unit{dBrad^2/Hz}\)) of the laser stabilized to a cryogenic silicon cavity \cite{Kedar-2023}.  Such phase noise level is calculated from \(S_\mathsf{y}(1\:\unit{Hz})=1.5{\times}10^{-33}~\unit{Hz^{-1}}\) \cite{Kedar-2023} using \(S_\varphi(f)=(\nu_0^2/f^2)\,S_\mathsf{y}(f)\).}
  \label{fig:PN_BOA_LASER_BENCH}
\end{figure*}

\section{First experiment}\label{sec:Experiment1}
The first experiment aims at validating the system in a reference condition, and at assessing the limits of what can be measured.  We do this in three steps, detailed below, with 150~\microW\ at the photodetector input.  Fig.~\ref{fig:PN_BOA_LASER_BENCH} shows the results.  

First, we measure the background noise of the system by symmetrizing the interferometer, with no DUT\@.  There is \(\approx\:20~\unit{m}\) fiber length in each arm, with \(\tau_D=380~\unit{ps}\) residual asymmetry (76~mm fiber length, group velocity \(2{\times}10^8~\unit{m/s}\)).  This ensures a large rejection of the laser noise.

\added{The background (Curve (1) on Fig.~\ref{fig:PN_BOA_LASER_BENCH}) is approximated with \(1.6{\times}10^{-5}/f\) \unit{rad^2/Hz} at \(f<45\)~Hz, \(3.2{\times}10^{-2}/f^3\) \unit{rad^2/Hz} from \(45~\unit{Hz}\) to \(5~\unit{kHz}\), and \(2.5{\times}10^{-13}\) \unit{rad^2/Hz} beyond 5~kHz.  That is, \(-48\), \(-15\) and \(-126\) \unit{dBrad^2/Hz} extrapolated to \(f=1~\unit{Hz}\), respectively.}  \added{It is rather irregular from 10~Hz to 1~kHz, and clean and smooth beyond}.
\added{The presence of \(1/f\) noise at low \(f\) is quite usual in differential measurements, where the delay cannot diverge in a finite measurement time, however long.  The irregularities from 10~Hz to 1~kHz point to technical noise.  
On the other hand, the \(1/f\) and \(1/f^3\) combo suggests that the background noise may be affected by the laser RIN, most likely with respect to its effect in the photodetector.  More in detail, this pattern is often encountered in the amplitude-noise spectrum \(S_\alpha(f)\) of oscillators, which is equivalent to RIN in optics because the fractional power fluctuations of \([1+\alpha(t)]\cos(2\pi\nu_0t)\) is \(2\alpha(t)\).  The full theory of AM noise is derived in \cite[Sec.\,7]{Rubiola-2010-GLE}.  The pattern we refer to is `Type~3' in Fig.\,23 but for the \(f>f_L\) region masked by the white noise.  However, the \(1/f^3\) region seems too wide to be ascribed to pollution from RIN.}

\added{Indeed, the most important fact is that the background noise is quite low.  Integrating Curve (1) in the full span of Fig.~\ref{fig:PN_BOA_LASER_BENCH} (1~Hz to 1~MHz) we find an RMS fluctuation of 7.8~mrad, mostly due to the \(1/f\) noise in the left-hand side of the plot, with a mere contribution of 0.1~mrad (1.3~\%) from the \(f>100\)~Hz region.  This is equivalent to a length fluctuation of 1.3~nm, or to a time fluctuation of \(6.4{\times}10^{-18}~\unit{s}\) in the optical system.  Such low background is achieved thanks to the differential measurement, where we measure \(\varphi(t)\) as the input-to-output phase of the DUT, rejecting the phase noise of the laser and of the RF source.}

\deleted{The background noise shows an irregular shape for \(f<1\) \unit{kHz}, and a clean flat white region \(S_\psi=2.5{\times}10^{-13}\) \unit{rad^2/Hz} (\(-126\) \unit{dBrad^2/Hz})}.

Second, we measure the laser noise \(S_\theta(f)\) with \(\tau_D=460\)~ns delay asymmetry (90~m optical fiber) in lieu of the DUT \added{(Curve (2) on Fig.~\ref{fig:PN_BOA_LASER_BENCH})}.
\(S_\theta(f)\) largely exceeds the background \added{(Curve(3))}, calculated from the above, using \eqref{eqn:H2f} and \(\tau_D=460\)~ns.  Thus, this is a trusted result, and validates the rejection of the background at small \(\tau_D\).
\(S_\theta(f)\) is approximated by the polynomial law \(S_\theta(f)=6.3{\times}10^{-9}/f^4+10^6/f^3~\unit{rad^2/Hz}\), which indicates the presence of frequency random walk below 10~kHz, and frequency flicker beyond.  Such phenomena are ubiquitous in lasers, and in RF/microwave oscillators as well.  
Using the formulas of the Enrico's Chart \cite[Fig.\,3]{Rubiola-2023-Companion}, the corresponding Allan deviation is \(\sigma(\tau)=6.1{\times}10^{-12}+1.1{\times}10^{-9}\sqrt{\tau}\), with a corner at \(\tau=33~\micros\) where flicker equals random walk.

In addition, artifacts in excess on the \(1/f^4\) noise appear at 50--1000~Hz, ascribed to acoustic noise, and a large flat bump in the \(1/f^3\) region centered at 150 kHz, still unexplained.

Finally, we re-introduce the DUT, we symmetrize the interferometer (negligible \(\tau_D\)) and we measure the total noise \(S_\psi(f)\), DUT plus background.  \added{This is Curve (4) on Fig.~\ref{fig:PN_BOA_LASER_BENCH}.}

At low frequency (4~Hz to 1.5~kHz), \replaced{phase noise with and without DUT are so close to one another that the DUT noise cannot be divided from the background.  In this region, the experiment outcome is that the measured spectrum is an \emph{upper bound} of the DUT noise.}{the DUT phase cannot be measured, hidden by the background.  Such background noise is likely due to environmental sensitivity, despite the highly stable conditions inside our shielded chamber.}
That said, the DUT noise is well visible in the \replaced{narrow}{small} 1--4~Hz region, but this is not significant enough to identify \added{clearly} a trend.   We see on Fig.~\ref{fig:PN_BOA_LASER_BENCH} that \(S_\varphi(1\,\unit{Hz})=6.3{\times}10^{-4}\;\unit{rad^2/Hz}\) (\(-32\)\,\unit{dBrad^2/Hz}).  We can take the approximation \(S_\varphi(f)=6.3{\times}10^{-4}/f~\unit{rad^2/Hz}\) as an upper bound of the DUT phase noise, thus the flicker coefficient is \(\mathsf{b}_{-1}=6.3{\times}10^{-4}~\unit{rad^2}\).  Using the formulas of the Enrico's Chart \cite[Fig.\,3]{Rubiola-2023-Companion}, we convert this value into the Allan deviation \(\sigma_\mathsf{y}(\tau)=5.3{\times}10^{-17}/\tau\).  This is lower than the instability of most Fabry-Pérot cavity stabilized lasers at \(\tau=1~\unit{s}\).  The instability of some of such lasers \cite{Xie-2017b, Kedar-2023, Guo-2022} is reported on Fig.~\ref{fig:PN_BOA_LASER_BENCH} for comparison.

\section{Second experiment}\label{sec:Experiment2}
The second experiment aims at assessing the DUT white phase noise of \(S_\psi(f)\) in different conditions of input power (\(P_i=0.8\ldots15~\microW\)) and pump current (\(I_P=150\ldots600~\unit{mA}\)).  We underline that in this Section we consider only the \emph{white phase noise} \(S_\psi\) and \(S_\varphi\), thus `\((f)\)' is omitted.

\begin{figure}[t]   
  {\centering\includegraphics[width=1\linewidth]{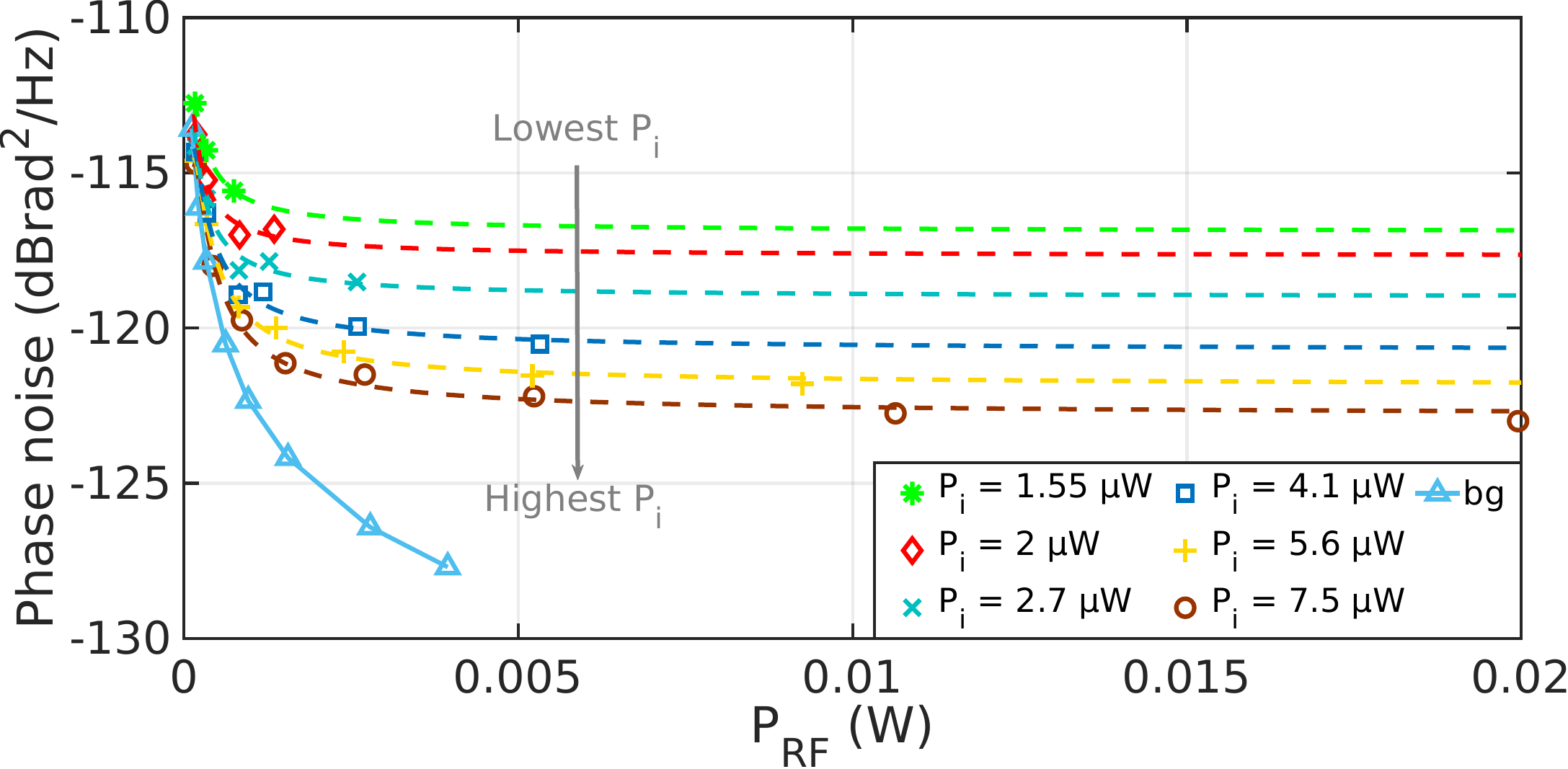}\par}
  \caption{White phase noise of the DUT measured at \(I_P=350~\unit{mA}\) pump power, for different values of the optical power \(P_i\) at the DUT input.  From the right hand to the left, the RF power is reduced by detuning the attenuator at the photodetector input.
  The background noise is shown for comparison.}
  \label{fig:PN_350mA}
\end{figure}

Figure~\ref{fig:PN_350mA} shows the results for \(I_P=350~\unit{mA}\)\@.
Notice that the horizontal axis is \(P_\text{RF}\), instead of \(P_0\).  The reason is that \(P_\text{RF}\) is measured directly, while \(P_0\) can only be estimated using \eqref{eqn:PRF}.
In fact, a power meter measures the total power, which is \(P_1+P_2+P_\text{ASE}\)\@.

The background noise of the photodiode and its RF amplifier is measured with the following trick.
\begin{enumerate}
  \item Symmetrize the power \(P_1=P_2=P_0\) of the two beams at the photodiode input by acting on the attenuator at the DUT input.  This is the best condition for low background noise.
  \item Set the DUT input power \(P_i\) to a convenient value.  \replaced{Higher \(P_i\) results in higher signal-to-noise ratio, but \(P_i\) must be kept well below the saturation level of the DUT and of the photodetector.  Keeping the amplifier in the linear region ensures that the symmetry condition is not affected by \(P_0\)}{This changes \(P_0\) without breaking the symmetry because the DUT is kept below the saturation point}.
  \item Attenuate progressively the optical beam at the photodiode input, until signal-to-noise ratio degrades.  This is better done by detuning the anti-ASE filter, rather than introducing an attenuator, because in this way the total ASE power---and its contribution to the background---is not affected.
\end{enumerate}
We observe that (i) \(S_\psi>S_{\psi\,\text{bg}}\), and that, (ii) close to the highest \(P_\text{RF}\), the degradation due to the attenuation is small enough.  In the end, we keep the right-hand point (highest \(P_\text{RF}\)) of each plot of Fig.~\ref{fig:PN_350mA} as the DUT noise \(S_\varphi\).  The other points (lower \(P_\text{RF}\)) serve only to validate the result. 

The experimental data are fitted with
\begin{align}
  S_\psi &= \frac{\alpha}{P_i^2} + \frac{\beta}{P_i} + \gamma \nonumber
\end{align}
where the values of \(\alpha\), \(\beta\) and \(\gamma\) are the outcome of the fit.  
We observe that \(\alpha\) and \(\beta\) match \eqref{eqn:PNbg2}, plus a possible small contribution from ASE, not further investigated.  Thus \(\gamma\) is the DUT phase noise \(S_\varphi\).

\begin{figure}[t]   
\includegraphics[width=1\linewidth]{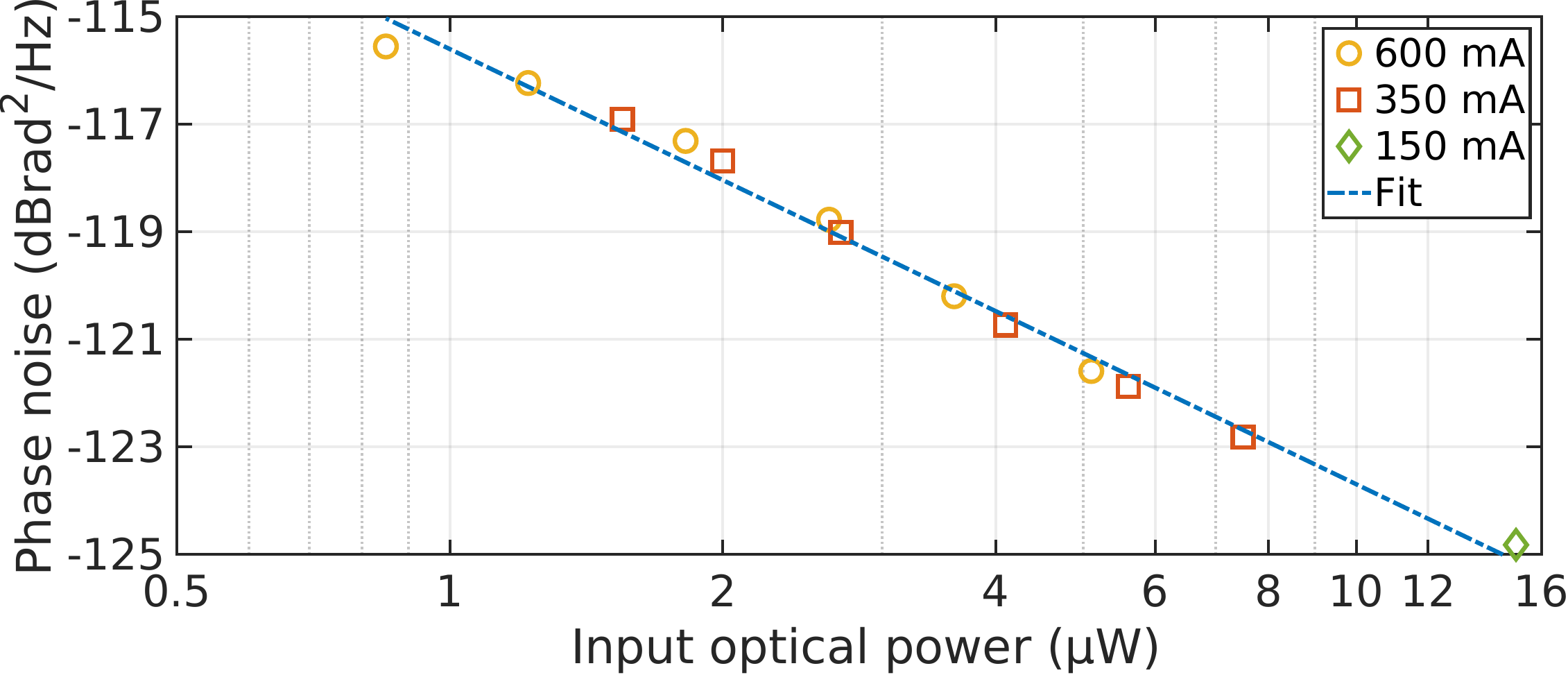}
\caption{Phase noise floor of the optical amplifier versus the SOA input optical power $P_i$, for several values (600, 350 and 150~mA) of the SOA driving current $I$. Only 1 point was taken for $I$ = 150 mA. In this case, the output RF signal was too low for lower values of $P_i$.}
\label{fig:PN_VS_PIN}
\end{figure}

The same measurement is repeated for 150~mA and 600~mA pump current.  A synthesis of the results is shown in Fig.~\ref{fig:PN_VS_PIN}.  The invalid points, \(S_\varphi\) hidden under the background, are omitted.  For this reason, at \(I_P=150~\unit{mA}\) we can show only one point.  Notice that the optical gain gets higher at higher \(I_P\), thus the experimental points are limited to lower \(P_i\).

Figure~\ref{fig:PN_VS_PIN} tells us that 
(i) no effect of the pump current \(I_P\) is observed, and 
(ii) the DUT white noise is \(S_\varphi\propto 1/P_i\), which is the phenomenological behavior of the shot noise.  The latter point matches the obvious facts that optical signals do not suffer from thermal noise, and shot noise is inherent in the stimulated emission.

\section{Conclusions}
We have discussed in detail a novel method for the phase-noise measurement of optical amplifiers using the delayed self-heterodyne intereferometric technique, and we have measured the phase noise of a semiconductor optical amplifier at 1.55~\micron\added{, operated below saturation}.  
\added{The method is suitable to laser amplifiers of other technologies, chiefly the EDFA, but the setup may need to be adapted (output power, gain, etc.)}.  

Taking \(-32~\unit{dBrad^2/Hz}\) at \(f=1~\unit{Hz}\) as an upper bound of the flicker noise, the amplifier noise limits the fractional frequency stability to \(5.2{\times}10^{-17}/\tau\) Allan deviation, as a function of the integration time \(\tau\).  This is lower than the instability of most Fabry-Pérot cavity-stabilized lasers \cite{Xie-2017b, Kedar-2023}. 
The flicker (\(1/f\)) phase noise could not be measured because it is of the same order or below the background.
Operating the optical amplifier under different conditions of input power \(P_i\) and pump current, we observed that the white phase noise is \added{(i)} independent of the pump current, and \replaced{(ii)}{it is} proportional to \(1/P_i\), as expected for shot noise, \deleted{and that it} down to \(-126\)~\unit{dBrad^2/Hz} at \(f\ge100\)~\unit{kHz}.

\section*{Funding}
This work is funded by Centre National d'Etudes Spatiales CNES (OSCAR project, Grant 200837/00), by the EIPHI Graduate school (Grant ANR-17-EURE-0002, REMICS sub-project), and by internal funds from Oscillator~IMP platform.

\section*{Data availability statement}
No database was generated during the current study, but the raw data are available from the corresponding author upon reasonable request.

%
\bibliographystyle{IEEEtran} 
\bibliography{PN-optical-amplifier}
\end{document}